\documentclass{article}
\usepackage{amsmath}
\usepackage{graphicx}

\begin{document}

\textbf{Trouton-Noble paradox revisited }\bigskip \bigskip

\qquad Tomislav Ivezi\'{c}

\qquad\textit{Ru%
\mbox
{\it{d}\hspace{-.15em}\rule[1.25ex]{.2em}{.04ex}\hspace{-.05em}}er Bo\v
{s}kovi\'{c} Institute, P.O.B. 180, 10002 Zagreb, Croatia}

\textit{\qquad ivezic@irb.hr\bigskip\bigskip}

\noindent An apparent paradox is obtained in all previous treatments of the
Trouton-Noble experiment; there is a three-dimensional (3D) torque $\mathbf{T%
}$ in an inertial frame $S$ in which a thin parallel-plate capacitor is
moving, but there is no 3D torque $\mathbf{T}^{\prime }$ in $S^{\prime }$,
the rest frame of the capacitor. Different explanations are offered for the
existence of another 3D torque, which is equal in magnitude but of opposite
direction giving that the total 3D torque is zero. In this paper instead of
using 3D quantities, \textit{e.g.}, $\mathbf{E}$, $\mathbf{B}$, $\mathbf{F}$%
, $\mathbf{L}$, $\mathbf{T}$, .. and their ``apparent'' transformations we
deal with 4D geometric quantities $E$, $B$, $K$, $M$, $N$, ... their Lorentz
transformations and equations with them. It is considered in our approach
that 4D geometric quantities and not the usual 3D quantities are
well-defined both theoretically and \emph{experimentally} in the 4D
spacetime. In analogy with the decomposition of the electromagnetic field $F$
(bivector) into two 1-vectors $E$ and $B$ we introduce decompositions of the
torque $N$ and the angular momentum $M$ (bivectors) into 1-vectors $N_{s}$, $%
N_{t}$ and $M_{s}$, $M_{t}$ respectively. The torques $N_{s}$, $N_{t}$ (the
angular momentums $M_{s}$, $M_{t}$), taken \emph{together, }contain the same
physical information as the bivector $N$ (the bivector $M$). It is shown
that in the frame of ``fiducial'' observers, in which the observers who
measure $N_{s}$ and $N_{t}$ are at rest, and in the standard basis, only the
spatial components $N_{s}^{i}$ and $N_{t}^{i}$ remain, which can be
associated with components of \emph{two} 3D torques $\mathbf{T}$ and $%
\mathbf{T}_{t}$. In such treatment with 4D geometric quantities the
mentioned paradox does not appear. The presented explanation is in a
complete agreement with the principle of relativity and with the
Trouton-Noble experiment without the introduction of any additional
torque.\bigskip \bigskip

\noindent \textbf{1. Introduction\bigskip }

\noindent In the experiment [1], see also [2], they looked for the turning
motion of a charged parallel plate capacitor suspended at rest in the frame
of the earth in order to measure the earth's motion through the ether. In
all previous treatments it is found that in the rest frame of a thin
parallel-plate capacitor, the $S^{\prime }$ frame, there is no
three-dimensional (3D) torque $\mathbf{T}$; $\mathbf{T}^{\prime }$ is zero
in $S^{\prime }$ since there is only an electric force between plates. (The
3D vectors will be designated in bold-face.) In the $S$ frame the capacitor
moves with uniform velocity $\mathbf{u}$ in the positive direction of the $%
x^{1}$ - axis. The charges on plates are now in uniform motion producing
both an electric field $\mathbf{E}$ and a magnetic field $\mathbf{B}$. The
existence of the magnetic field $\mathbf{B}$ in $S$ is responsible for the
existence of the 3D magnetic force and this force provides a 3D torque on
the charged capacitor. In that way an apparent paradox, the Trouton-Noble
paradox, is obtained and the principle of relativity is violated; there is a
3D torque and so a time rate of change of 3D angular momentum in one
inertial frame, but no 3D angular momentum and no 3D torque in another.

Different explanations have been offered for the existence of another 3D
torque which is equal in magnitude but of opposite direction giving that the
total 3D torque is zero in order to have the agreement with the principle of
relativity and experiments. All these previous explanations for the null
result of the experiments, see, \textit{e.g.}, [3-7] and references therein,
mainly deal with 3D quantities, \textit{e.g.}, $\mathbf{E}$, $\mathbf{B}$, $%
\mathbf{F}$, $\mathbf{L}$, $\mathbf{T}$, their ``apparent'' transformations
and equations with them.

However a new explanation of the Trouton-Noble experiment, which is
developed in [8], deals from the outset with 4D geometric quantities, their
Lorentz transformations (LT) and equations with them. The main point in such
geometric approach as in [8] is that the physical meaning, both
theoretically and \emph{experimentally}, is attributed to 4D geometric
quantities, and not, as usual, to 3D quantities. The consideration from [8]
reveals that there is no need either for the nonelectromagnetic forces and
their additional torque, [3-6], or for the angular electromagnetic field
momentum and its rate of change, i.e., its additional torque, [7].
Particularly the ``resolution'' of the Trouton-Noble paradox from [7] is
examined in more detail in [8]. There it is shown that, when $r^{\prime 0}=0$%
, where $r^{\prime 0}$ is the temporal component of the lever arm joining
the axis of rotation with the point of application of the resultant 3D
force, then the 4D torque $N$ (bivector) is zero not only in $S^{\prime }$,
the rest frame of a thin parallel-plate capacitor, but in all other
relatively moving inertial frames.

In this paper, sections 3-3.2, the resolution of the Trouton-Noble paradox
with the 4D torque $N$ will be given in the case when the temporal component
of the lever arm is zero, $r^{0}=0$, in the $S$ frame in which the capacitor
is moving. The case $r^{0}=0$ is not investigated in [8], but, in some
sense, it corresponds to the case investigated in [7]. Here it will be shown
that the torque $N$ as a 4D geometric quantity is the same quantity in both
frames $S^{\prime }$ and $S$, in contrast to the usual approaches, e.g.,
[7], in which the torque $\mathbf{T}^{\prime }$ in $S^{\prime }$ is zero, $%
\mathbf{T}^{\prime }=\mathbf{0}$, while it is different from zero in the $S$
frame, $\mathbf{T\neq 0.}$

Furthermore, section 3.1, we shall introduce the decomposition of the 4D
torque $N$ into two 1-vectors, the ``space-space'' torque $N_{s}$ and the
``time-space'' torque $N_{t}$, which is first presented in [9]. In the frame
of ``fiducial'' observers, in which the observers who measure $N_{s}$ and $%
N_{t}$ are at rest, and in the standard basis, only the spatial components $%
N_{s}^{i}$ and $N_{t}^{i}$ remain, which can be associated with components
of \emph{two} 3D torques $\mathbf{T}$ and $\mathbf{T}_{t}$. Note that in all
usual approaches, including [7], only the 3D torque $\mathbf{T}=\mathbf{r}%
\times \mathbf{F}$ is mentioned and considered as the physical one. The 4D
torques $N_{s}$ and $N_{t}$ will be calculated for both cases, $r^{\prime
0}=0$ and $r^{0}=0$. It will be shown in section 3.3 that in the approach
with the 4D torques $N_{s}$ and $N_{t}$, in the same way as in the approach
with the 4D torque $N$ from [8], the principle of relativity is naturally
satisfied and there is no paradox.

Some objections to the calculation of $\mathbf{T}$ in [7] are raised in
section 2, while the detailed comparison of our calculation of $N$, $N_{s}$
and $N_{t}$ and the usual calculation of $\mathbf{T}$ is given in section 4.
In section 5 the conclusions are presented. \bigskip \bigskip

\noindent \textbf{2.}\textit{\ }\textbf{The choice of the basis in the 4D
spacetime and some remarks }

\textbf{on the calculation of the 3D torque }$\mathbf{T}$ \textbf{from [7] }%
\bigskip

\noindent The whole investigation will be done in the geometric algebra
formalism, see, e.g., [10,11]. Physical quantities will be represented by 4D
geometric quantities, multivectors, that are defined without reference
frames, i.e., as absolute quantities (AQs) or, when some basis has been
introduced, these AQs are represented as 4D coordinate-based geometric
quantities (CBGQs) comprising both components and a basis. For simplicity
and for easier understanding, only the standard basis \{$\gamma _{\mu };\
0,1,2,3$\} of orthonormal 1-vectors, with timelike vector $\gamma _{0}$ in
the forward light cone, will be used, but remembering that the approach with
4D geometric quantities holds for any choice of basis.

It is worth noting that the standard basis $\left\{ \gamma _{\mu }\right\} $
corresponds, in fact, to the Einstein's system of coordinates. In Einstein's
system of coordinates the standard, i.e., Einstein's synchronization [12] of
distant clocks and Cartesian space coordinates $x^{i}$ are used in the
chosen inertial frame. However different systems of coordinates of an
inertial frame are allowed and they are all equivalent in the description of
physical phenomena. For example, in [13,14] and in the second and the third
paper in [15], two very different, but physically completely equivalent
systems of coordinates, Einstein's system of coordinates and the system of
coordinates with a nonstandard synchronization, the everyday (radio) (``r'')
synchronization, are exposed and exploited throughout the papers.

We shall again, as in [8], examine the situation considered in [7].
Regarding the figures in [7] it has to be remarked that both figures would
need to contain the time axes as well. Figure 1 (figure 2) is the projection
onto the hypersurface $t^{\prime }=0$ ($t=0$); the distances are
simultaneously determined in the $S^{\prime }$ ($S$) frame. This means that
figure 1 corresponds to the choice $r^{\prime 0}=0$ and figure 2 to the
choice $r^{0}=0$. The LT cannot transform the hypersurface $t^{\prime }=0$
into the hypersurface $t=0$ and the distances that are simultaneously
determined in the $S^{\prime }$ frame cannot be transformed by the LT into
the distances simultaneously determined in the $S$ frame. Such figures could
be possible for the Galilean transformations, but for the LT they are
meaningless.

The transformations for components of the 3D torque $\mathbf{T}$, equations
(1)-(3) in [7], are found, e.g., in Jefimenko's book [16], equations
(8-6.11)-(8-6.13). They are

\begin{equation}
T_{1}=T_{1}^{\prime }/\gamma ,\ T_{2}=T_{2}^{\prime }+\beta
^{2}r_{1}^{\prime }F_{3}^{\prime },\ T_{3}=T_{3}^{\prime }-\beta
^{2}r_{1}^{\prime }F_{2}^{\prime },  \label{s}
\end{equation}
where $\beta =\left| \mathbf{u}\right| /c$, $\gamma =(1-\left| \mathbf{u}%
\right| ^{2}/c^{2})^{-1/2}$. In [16] these transformations are derived
starting with the definition of the 3D torque $\mathbf{T}=\mathbf{r}\times
\mathbf{F}$ and using the LT, but only for the spatial components $r_{i}$,
i.e., $r_{x}$, $r_{y}$, $r_{z}$, and the transformations for components of
the 3D force $\mathbf{F}$, equations (8.5.1)-(8.5.3) in [16]. Observe that $%
t=0$ is chosen to be the time of observation in $S$. Actually what is
assumed in that derivation from [16] is not $t=0$, but that $r^{0}=0$. Then
the transformation for the time component is $r^{0}=\gamma (r^{\prime
0}+\beta r^{\prime 1})=0$, which yields $r^{\prime 0}=-\beta r^{\prime 1}$.
That relation is used in the derivation of equations (8-6.11)-(8-6.13) in
[16], i.e., (\ref{s}). Physically $r^{0}=0$ means that the lever arm is
simultaneously determined in the $S$ frame in which the capacitor is moving.
In the Trouton-Noble experiment the $S$ frame refers to the preferred frame,
while the $S^{\prime }$ frame refers to the Earth. Hence the appropriate
choice for the comparison with experiment would be that $r^{\prime 0}=0$,
i.e., that the lever arm is simultaneously determined in its rest frame, the
$S^{\prime }$ frame. It is worth noting that in [7] Jefimenko deals only
with the 3D quantities thus implicitly assuming that the lengths, volumes
and angles are well defined in \emph{both} relatively moving inertial
frames, see figures 1 and 2 and, e.g., equation (10) in [7]. Since the
relations (\ref{s}) are obtained for the case $r^{0}=0$, then the mentioned
3D quantities are well-defined only in the $S$ frame, but not in the $%
S^{\prime }$ frame. In the 4D spacetime it is not possible to have that,
e.g., the lengths, are well defined in \emph{both} relatively moving
inertial frames.

In order to avoid such ambiguities we have calculated in [8] the 4D torque $%
N $ and showed that there is no Trouton-Noble paradox for the case $%
r^{\prime 0}=0$, when the 4D geometric quantities are used. From the
experimental point of view the other case, $r^{0}=0$, is difficult to
realize in measurements. Regardless of that in this paper we shall
investigate the case $r^{0}=0$ as well in order to see the fundamental
difference between the usual approaches, e.g., [7] and [16], and our
approach in which the physical reality is attributed to the 4D geometric
quantities.\bigskip \bigskip

\noindent \textbf{3.}\textit{\ }\textbf{The resolution of the Trouton-Noble
paradox representing }

\textbf{the torque by the bivector }$N$ \textbf{and by the 1-vectors }$N_{s}$
\textbf{and} $N_{t}$ \bigskip

\noindent The discussion and the results presented in [8] and [9] strongly
suggest that the relativistically correct resolution of the Trouton-Noble
paradox can be achieved in an unambiguous way by the use of the 4D torque $N$%
, or $N_{s}$ and $N_{t}$, and not with the usual 3D torque $\mathbf{T}$.
This will be realized in sections 3.1-3.3. \bigskip \medskip

\noindent \textit{3.1. The torques }$N$, $N_{s}$ and $N_{t}$ \textit{%
\bigskip }

\noindent As shown in [8] the torque $N$, as a 4D AQ, is defined as a
bivector $N=r\wedge K$,$\ $where $r=x_{P}-x_{O}$. $r$ is 1-vector associated
with the lever arm, $x_{P}$ and $x_{O}$ are the position 1-vectors
associated with the spatial point of the axis of rotation and the spatial
point of application of the force $K$. $P$ and $O$ are the events whose
position 1-vectors are $x_{P}$ and $x_{O}$. We shal need to determine $N$
for the Lorentz force $K_{L}$ which is $K_{L}=(q/c)F\cdot u$, where $u$ is
the velocity 1-vector and $F$ is the electromagnetic field $F(x)$
(bivector). This expression for $K_{L}$ with $F$ is used in [8], but in this
paper the decomposition of $F$ into electric and magnetic fields will be
employed.

It is proved in the tensor formalism that given an antisymmeric tensor (as
geometric quantity) $F^{ab}$ and a unit time-like four-vector $n^{a}$ one
can construct two four-vectors, $E^{a}=F^{ab}n_{b}$, and $%
B^{a}=(1/2)\varepsilon ^{abcd}F_{cd}n_{b}$ and, oppositely, that $F^{ab}$
can be expressed in terms of these two four-vectors $E^{a}$, $B^{a}$ and $%
n^{a}$ as $F^{ab}=E^{a}n^{b}-E^{b}n^{a}+\varepsilon ^{abcd}B_{c}n_{d}$, see,
e.g., [17], section 6, Example 6.1 and [18], or in the covariant form [19],
equation (7.58). This decomposition of an antisymmetric tensor is used for
the decomposition of the tensor of the electromagnetic field $F^{ab}$ in
[13] [20] [14] and [21]. The same decomposition but in the Clifford algebra
formalism is introduced and employed in [22-24] and [9]. The electromagnetic
field $F$ (bivector) is decomposed into 1-vectors of the electric field $E$
and the magnetic field $B$ and a unit time-like 1-vector $v/c$ as
\begin{align}
F& =(1/c)E\wedge v+(IB)\cdot v,  \notag \\
E& =(1/c)F\cdot v,\quad B=-(1/c^{2})I(F\wedge v),  \label{itf}
\end{align}
where $I$ is the unit pseudoscalar and $v$ is the velocity (1-vector) of a
family of observers who measures $E$ and $B$ fields. It also holds that $%
E\cdot v=B\cdot v=0$, which yields that only three components of $E$ and
three components of $B$ are independent quantities. Observe that $E$ and $B$
depend not only on $F$ but on $v$ as well.

Using the decomposition of $F$ into $E$ and $B$, (\ref{itf}), the Lorentz
force $K_{L}$ can be written as $K_{L}=(q/c)F\cdot u=(q/c)\left[
(1/c)E\wedge v+(IB)\cdot v\right] \cdot u$, where $u$ is the velocity
(1-vector) of a charge $q$, [22-24]. Particularly, from the definition of
the Lorentz force $K_{L}=(q/c)F\cdot u$ and the relation $E=(1/c)F\cdot v$
it follows that the Lorentz force ascribed by an observer comoving with a
charge, $u=v$, is purely electric $K_{L}=qE$.

The 4D torque $N=r\wedge K_{L}\ $can be represented as a 4D CBGQ. In $%
S^{\prime }$, the rest frame of the capacitor, it becomes $%
N=(1/2)N^{^{\prime }\mu \nu }\gamma _{\mu }^{\prime }\wedge \gamma _{\nu
}^{\prime }$, $N^{\prime \mu \nu }=r^{\prime \mu }K_{L}^{\prime \nu
}-r^{\prime \nu }K_{L}^{\prime \mu }$, where the components $N^{\prime \mu
\nu }$ are determined as $N^{\prime \mu \nu }=\gamma ^{\prime \nu }\cdot
(\gamma ^{\prime \mu }\cdot N)$.

In [8] only the 4D torque $N$ is considered. Here we shall introduce new 4D
torques, 1-vectors $N_{s}$ and $N_{t}$. In fact, they are first introduced
in [9]. There the same decomposition as for $F$, (\ref{itf}), is made for
the 4D torque $N$; it is decomposed into two 1-vectors, the ``space-space''
torque $N_{s}$ and the ``time-space'' torque $N_{t}$, and the unit time-like
1-vector $v/c$ as
\begin{eqnarray}
N &=&(v/c)\cdot (IN_{s})+(v/c)\wedge N_{t}  \notag \\
N_{s} &=&I(N\wedge v/c),\quad N_{t}=(v/c)\cdot N,  \label{nls}
\end{eqnarray}
with the condition
\begin{equation}
N_{s}\cdot v=N_{t}\cdot v=0;  \label{cs}
\end{equation}
only three components of $N_{s}$ and three components of $N_{t}$ are
independent since $N$ is antisymmetric. Here again $v$ is the velocity
(1-vector) of a family of observers who measures $N_{s}$ and $N_{t}$.
Similarly as for $E$ and $B$ the 4D torques $N_{s}$ and $N_{t}$ depend not
only on the bivector $N$ but on $v$ as well. The relations (\ref{nls}) show
that $N_{s}$ and $N_{t}$ taken\emph{\ together} contain the same physical
information as the bivector $N$.

When $N_{s}$ and $N_{t}$ are written as CBGQs in the $\{\gamma _{\mu
}^{\prime }\}$ basis they are
\begin{equation}
N_{s}=N_{s}^{\prime \mu }\gamma _{\mu }^{\prime }=(1/2c)\varepsilon ^{\alpha
\beta \mu \nu }N_{\alpha \beta }^{\prime }v_{\mu }^{\prime }\gamma _{\nu
}^{\prime },\ N_{t}^{\prime }=(1/c)N^{\prime \mu \nu }v_{\mu }^{\prime
}\gamma _{\nu }^{\prime }.  \label{lg}
\end{equation}
Let us take that the $S^{\prime }$ frame is the frame of ``fiducial''
observers, or the $\gamma _{0}^{\prime }$ - frame, in which the observers
who measure 1-vectors $E$, $B$, $K_{L}$, $N_{s}$ and $N_{t}$ are at rest.
Then in $S^{\prime }$ the velocity $v$ is $v=c\gamma _{0}^{\prime }$. In the
$\gamma _{0}^{\prime }$ - frame and the $\{\gamma _{\mu }^{\prime }\}$ basis
$v$ has the components $v^{\prime \mu }=(c,0,0,0)$. Hence in the frame of
``fiducial'' observers (\ref{lg}) becomes
\begin{equation}
N_{s}^{\prime 0}=0,N_{s}^{\prime i}=(1/2)\varepsilon ^{0jki}N_{jk}^{\prime
},\ N_{t}^{\prime 0}=0,\ N_{t}^{\prime i}=N^{\prime 0i}.  \label{l1}
\end{equation}
It is seen from (\ref{l1}) that $N_{s}^{\prime 0}=N_{t}^{\prime 0}=0$ and
only the spatial components remain. $N_{s}^{\prime i}$ components are $%
N_{s}^{\prime 1}=N^{\prime 23}=r^{\prime 2}K_{L}^{\prime 3}-r^{\prime
3}K_{L}^{\prime 2}$, $N_{s}^{\prime 2}=N^{\prime 31}$ and $N_{s}^{\prime
3}=N^{\prime 12}$.

The 1-vector $N_{s}$ corresponds to the 3D torque $\mathbf{T}$ that is
considered in [7] and [16]. On the other hand in [7] and [16], and in all
other approaches that deal with the 3D quantities, there is no 3D torque
which would correspond to our 1-vector $N_{t}$. Nevertheless we shall
introduce already here another 3D torque $\mathbf{T}_{t}$, which corresponds
to our 4D torque $N_{t}$. The precise meaning of the mentioned
correspondence and of the 3D torque $\mathbf{T}_{t}$ will be better
explained later.

The whole discussion with the torque can be completely repeated for the
angular momentum replacing $N$, $N_{s}$ and $N_{t}$ by $M$, $M_{s}$ and $%
M_{t}$, see [9]. The angular momentum $M$ as a 4D AQ (bivector) and
manifestly Lorentz invariant equation connecting $M$ and $N$ are defined as $%
M=r\wedge p$, $N=dM/d\tau $, where $p$ is the proper momentum (1-vector) and
$\tau $ is the proper time. The 1-vectors $M_{s}$ and $M_{t}$ correspond to $%
\mathbf{L}$ and $\mathbf{L}_{t}$ respectively in the usual 3D picture. $%
\mathbf{L}$ and $\mathbf{L}_{t}$ are introduced in [25]. The components $%
L_{i}$ of the 3D vector $\mathbf{L}$ (which is called the angular momentum)
are identified with the ``space-space'' components of the covariant angular
momentum tensor $M^{\mu \nu }$, $M^{\mu \nu }=x^{\mu }p^{\nu }-x^{\nu
}p^{\mu }$, and the components $L_{t,i}$ of the 3D vector $\mathbf{L}_{t}$
(for which a physical interpretation is not given in [25]) are identified
with the three ``time-space'' components of $M^{\mu \nu }$. (We denote
Jackson's $K_{i}$ with $L_{t,i}$, $\mathbf{K}$ with $\mathbf{L}_{t}$.)

In the usual picture with 3D quantities one can connect $\mathbf{T}$ and $%
\mathbf{T}_{t}$ with the above mentioned $\mathbf{L}$ and $\mathbf{L}_{t}$
by the relations $\mathbf{T}=d\mathbf{L}/dt$ and $\mathbf{T}_{t}=d\mathbf{L}%
_{t}/dt$.

As shown in [8] all ``space-space'' components $N^{\prime ij}$ and the
``time-space'' components $N^{\prime 0i}$ are zero, $N^{\prime \mu \nu }=0$,
in $S^{\prime }$ and in the case when $r^{\prime 0}=0$. Accordingly, in $%
S^{\prime }$, the whole bivector $N$ is zero, $N=0$, when $r^{\prime 0}=0$.
As already mentioned the case $r^{\prime 0}=0$ is the appropriate choice for
the comparison with experiment since the lever arm is simultaneously
determined in $S^{\prime }$ in which the capacitor is at rest. The essential
difference between our geometric approach and the usual covariant picture is
the presence of the basis. The existence of the basis causes that every 4D
CBGQ is invariant under the passive LT; the components transform by the LT
and the basis by the inverse LT leaving the whole 4D CBGQ unchanged. This
means that a CBGQ represents \emph{the same physical quantity }for
relatively moving 4D observers. For the torque $N=(1/2)N^{\prime \mu \nu
}\gamma _{\mu }^{\prime }\wedge \gamma _{\nu }^{\prime }$ this yields that
the components transform by the LT as
\begin{eqnarray}
N^{23} &=&N^{\prime 23},\ N^{31}=\gamma (N^{\prime 31}-\beta N^{\prime
03}),\ N^{12}=\gamma (N^{\prime 12}+\beta N^{\prime 02}),  \notag \\
N^{01} &=&N^{\prime 01},\ N^{02}=\gamma (N^{\prime 02}+\beta N^{\prime
12}),\ N^{03}=\gamma (N^{\prime 03}+\beta N^{\prime 13}),  \label{nc}
\end{eqnarray}
whereas the basis $\gamma _{\mu }^{\prime }\wedge \gamma _{\nu }^{\prime }$
transform by the inverse LT giving that the whole 4D torque $N$ is unchanged
\begin{equation}
N=(1/2)N^{^{\prime }\mu \nu }\gamma _{\mu }^{\prime }\wedge \gamma _{\nu
}^{\prime }=(1/2)N^{\mu \nu }\gamma _{\mu }\wedge \gamma _{\nu }.
\label{enc}
\end{equation}
From (\ref{enc}) it is concluded in [8] that the whole 4D torque $N$ is zero
not only in the rest frame of the capacitor but in all other relatively
moving inertial frames of reference. Thus it is proved in [8] that for $%
r^{\prime 0}=0$ the principle of relativity is naturally satisfied and there
is no Trouton-Noble paradox.

It is immediately seen from the definitions of $N_{s}$ and $N_{t}$, (\ref
{nls}), that they are also zero in all relatively moving inertial frames of
reference when $r^{\prime 0}=0,$
\begin{equation}
r^{\prime 0}=0\Rightarrow N_{s}=N_{t}=0.  \label{st}
\end{equation}
Again we conclude that there is no Trouton-Noble paradox.\medskip \bigskip

\noindent 3\textit{.2. The 4D torque }$N$ \textit{when }$r^{0}=0$ \textit{%
\bigskip }

\noindent Let us now consider the case when $r^{\prime 0}\neq 0$, but $%
r^{0}=0$ ($r^{\prime 0}=-\beta r^{\prime 1}$). This case is not investigated
in [8]. As already mentioned the case $r^{0}=0$ corresponds to that one
explored in [7] section 2 and [16]. In the $S^{\prime }$ frame, which is
again chosen to be the frame of ``fiducial'' observers, the velocity $u$ of
the capacitor is $u=c\gamma _{0}^{\prime }$, or $u=u^{\prime \mu }\gamma
_{\mu }^{\prime }$, where $u^{\prime \mu }=(c,0,0,0)$. The components $%
r^{\prime \mu }$ are $r^{\prime \mu }=(r^{\prime 0},r^{\prime 1},r^{\prime
2},0)$ where $r^{\prime 1}=r_{x}^{\prime }=-a^{\prime }\sin \Theta ^{\prime
},r^{\prime 2}=r_{y}^{\prime }=a^{\prime }\cos \Theta ^{\prime },r^{\prime
3}=r_{z}^{\prime }=0$, see figure 1 in [7]. Hence in $S^{\prime }$ $%
v=u=c\gamma _{0}^{\prime }$. As already mentioned in that case the Lorentz
force is purely electric, $K_{L}=qE$, where $q$ is the total charge residing
on the positive plate. Then $K_{L}^{\prime \nu }=(0,qE^{\prime 1},qE^{\prime
2},0)$ or $K_{L}^{\prime \nu }=(0,F_{x}^{\prime },F_{y}^{\prime },0)$, where
$F_{x}^{\prime }$ and $F_{y}^{\prime }$ are the components of the 3D force $%
\mathbf{F}$ that are the same as in [7], but written with primed quantities;
$F_{x}^{\prime }=Cr_{x}^{\prime }$, $F_{y}^{\prime }=Cr_{y}^{\prime }$, $%
F_{z}^{\prime }=0$ and $C=-\sigma ^{\prime 2}A^{\prime }/2\varepsilon
_{0}a^{\prime }$. This yields that the components $N^{^{\prime }\mu \nu }$
are
\begin{eqnarray}
N^{\prime 12} &=&N^{\prime 13}=N^{\prime 23}=0,  \notag \\
N^{\prime 01} &=&r^{\prime 0}K_{L}^{\prime 1}-r^{\prime 1}K_{L}^{\prime
0}=r^{\prime 0}F_{x}^{\prime },\ N^{\prime 02}=r^{\prime 0}F_{y}^{\prime },\
N^{\prime 03}=0.  \label{c1}
\end{eqnarray}
where the result that $N^{\prime 12}=r^{\prime 1}K_{L}^{\prime 2}-r^{\prime
2}K_{L}^{\prime 1}=r_{x}^{\prime }F_{y}^{\prime }-r_{y}^{\prime
}F_{x}^{\prime }=0$ is obtained inserting the explicit expressions for $%
r_{x}^{\prime }$, $r_{y}^{\prime }$, and $F_{x}^{\prime }$, $F_{y}^{\prime }$
into $N^{\prime 12}$. Hence in $S^{\prime }$ the whole torque $N$ is given
as
\begin{eqnarray}
r^{0} &=&0,\ N=(1/2)N^{\prime \mu \nu }\gamma _{\mu }^{\prime }\wedge \gamma
_{\nu }^{\prime }=N^{\prime 01}\gamma _{0}^{\prime }\wedge \gamma
_{1}^{\prime }+N^{\prime 02}\gamma _{0}^{\prime }\wedge \gamma _{2}^{\prime
},  \notag \\
N^{\prime 01} &=&-\beta r_{x}^{\prime }F_{x}^{\prime },\quad N^{\prime
02}=-\beta r_{x}^{\prime }F_{y}^{\prime }.  \label{gi}
\end{eqnarray}

In the $S$ frame $N=(1/2)N^{\mu \nu }\gamma _{\mu }\wedge \gamma _{\nu }$
and the components $N^{\mu \nu }$ can be determined writing $N^{\mu \nu
}=r^{\mu }K_{L}^{\nu }-r^{\nu }K_{L}^{\mu }$ and using the LT for $r^{\mu }$
and $K_{L}^{\mu }$, or using directly the LT for $N^{\mu \nu }$, which are
given by (\ref{nc}). Taking into account that in $S^{\prime }$ all $%
N^{\prime ij}=0$ we find that the ``space-space'' components $N^{ij}$ are $%
N^{12}=\gamma \beta N^{\prime 02}$ and $N^{13}=N^{23}=0$. From the above
relations we see that the ``time-space'' components $N^{0i}$ are the
following, $N^{01}\neq 0$, $N^{02}\neq 0$ and $N^{03}=0$, which yields that
the whole $N$ in $S$ is
\begin{eqnarray}
r^{0} &=&0,\ N=(1/2)N^{\mu \nu }\gamma _{\mu }\wedge \gamma _{\nu
}=N^{01}\gamma _{0}\wedge \gamma _{1}+N^{02}\gamma _{0}\wedge \gamma
_{2}+N^{12}\gamma _{1}\wedge \gamma _{2},  \notag \\
N^{01} &=&-\beta r_{x}^{\prime }F_{x}^{\prime },\quad N^{02}=-\gamma \beta
r_{x}^{\prime }F_{y}^{\prime },\quad N^{12}=-\gamma \beta ^{2}r_{x}^{\prime
}F_{y}^{\prime }.\   \label{g1}
\end{eqnarray}
Consequently, when $r^{0}=0$, we find that in the $S$ frame, in which the
capacitor is moving, the components $N^{\mu \nu }$ that are different from
zero are not only the ``time-space'' components $N^{01}$ and $N^{02}$, but
also the ``space-space'' component $N^{12}$.

Of course, due to the invariance of any CBGQ under the passive LT, $N$ from (%
\ref{g1}) is equal to $N$ from (\ref{gi}),
\begin{equation}
r^{0}=0,\ N(\ref{gi})=N(\ref{g1}).  \label{jd}
\end{equation}
This means that the principle of relativity is again naturally satisfied and
there is no Trouton-Noble paradox for the whole 4D torque $N$.

It is interesting to explore the connection between our result for the whole
4D torque $N$ (\ref{g1}) and the usual result for the 3D torque $\mathbf{T}$%
\textbf{\ }from [7]. From (\ref{nc}) we know that $N^{12}=\gamma N^{\prime
12}+\gamma \beta N^{\prime 02}=\gamma \lbrack (r^{\prime 1}K_{L}^{\prime
2}-r^{\prime 2}K_{L}^{\prime 1})+\beta (r^{\prime 0}K_{L}^{\prime
2}-r^{\prime 2}K_{L}^{\prime 0})]$. Since $S^{\prime }$ is the frame of
``fiducial'' observers and at the same time it is the rest frame of the
capacitor, $v=u=c\gamma _{0}^{\prime }$, one can write $N^{12}$ in terms of
the components of the 3D force $\mathbf{F}$, which yields $N^{12}=\gamma
(r_{x}^{\prime }F_{y}^{\prime }-r_{y}^{\prime }F_{x}^{\prime })+\gamma \beta
\lbrack (-\beta r_{x}^{\prime })F_{y}^{\prime }]$. Then all ``space-space''
components $N^{ij}$ can be written in terms of the components of the 3D
torque $\mathbf{T}$ as
\begin{equation}
N^{12}=\gamma (T_{z}^{\prime }-\beta ^{2}r_{x}^{\prime }F_{y}^{\prime }),\
N^{31}=\gamma (T_{y}^{\prime }+\beta ^{2}r_{x}^{\prime }F_{z}^{\prime }),\
N^{23}=T_{x}^{\prime },  \label{ss}
\end{equation}
When it is taken that all $N^{\prime ij}=0$ and that $F_{z}^{\prime }=0$
then $N^{23}=N^{31}=0$ and only $N^{12}$ remains, $N^{12}=\gamma (-\beta
^{2}r_{x}^{\prime }F_{y}^{\prime })$, as in (\ref{g1}). This component $%
N^{12}$ corresponds to the component $T_{z}$, equation (4) in [7]. However,
it is worth noting that in all usual approaches, including [7], only three
components of the 3D torque $\mathbf{T}$\textbf{\ }are considered to be
physical. In our approach physically important, measurable, quantity is the
whole 4D torque $N$. When that $N$ is written as a CBGQ then it contains
sixteen components $N^{\mu \nu }$ (six of them are independent) together
with the bivector basis $\gamma _{\mu }\wedge \gamma _{\nu }$ and equation (%
\ref{enc}) holds.\medskip \bigskip

\noindent \textit{3.3. The 4D torques }$N_{s}$ and $N_{t}$ \textit{when }$%
r^{0}=0$ \textit{\bigskip }

\noindent Let us now calculate the torques $N_{s}$ and $N_{t}$ in the case
when $r^{0}=0$. Again we assume that $S^{\prime }$ is the frame of
``fiducial'' observers, i.e., $v=u=c\gamma _{0}^{\prime }$. Then, as seen
from (\ref{gi}), all ``space-space'' components $N^{\prime ij}$ are zero.
Using (\ref{lg}), or (\ref{l1}), one finds that in our case all $%
N_{s}^{\prime \mu }=0$ in $S^{\prime }$. According to that the whole $N_{s}$
is zero, $N_{s}=N_{s}^{\prime \mu }\gamma _{\mu }^{\prime }=0$, when $%
r^{0}=0 $. From (\ref{l1}) it follows that $N_{t}^{\prime 0}=0$, $%
N_{t}^{\prime i}=N^{\prime 0i}$, where $N^{\prime 0i}=r^{\prime
0}K_{L}^{\prime i}=-\beta r^{\prime 1}K_{L}^{\prime i}$. Hence we have
\begin{eqnarray}
r^{0} &=&0,\ N_{s}=N_{s}^{\prime \mu }\gamma _{\mu }^{\prime }=0,\quad
N_{t}=N_{t}^{\prime \mu }\gamma _{\mu }^{\prime },  \notag \\
N_{t}^{\prime 0} &=&0,\ N_{t}^{\prime 1}=-\beta r_{x}^{\prime }F_{x}^{\prime
},\ N_{t}^{\prime 2}=-\beta r_{x}^{\prime }F_{y}^{\prime },\ N_{t}^{\prime
3}=0.  \label{s2}
\end{eqnarray}

Next we determine $N_{s}$ and $N_{t}$ in the $S$ frame. Note that, e.g., $%
N_{s}^{\prime \mu }\gamma _{\mu }^{\prime }$ transforms under the LT as
every 1-vector transforms, which means that components $N_{s}^{\prime \mu }$
(of the ``space-space'' torque $N_{s}$) transform to the components $%
N_{s}^{\mu }$ (of the same torque $N_{s}$ in the $S$ frame); there is no
mixing with the components of the ``time-space'' torque $N_{t}$
\begin{equation}
N_{s}^{0}=\gamma (N_{s}^{\prime 0}+\beta N_{s}^{\prime 1}),\
N_{s}^{1}=\gamma (N_{s}^{\prime 1}+\beta N_{s}^{\prime 0}),\
N_{s}^{2,3}=N_{s}^{\prime 2,3}.  \label{aa}
\end{equation}
The unit 1-vectors $\gamma _{\mu }^{\prime }$ transform by the inverse LT.
Only with such transformations, e.g., the CBGQs $N_{s}^{\prime \mu }\gamma
_{\mu }^{\prime }$ and $N_{s}^{\mu }\gamma _{\mu }$ are the same quantity $%
N_{s}$ in $S^{\prime }$ and $S$ frames, $N_{s}=N_{s}^{\prime \mu }\gamma
_{\mu }^{\prime }=N_{s}^{\mu }\gamma _{\mu }$. The same is fulfilled for $%
N_{t}^{\prime \mu }\gamma _{\mu }^{\prime }$. Thus it holds that in the $S$
frame too $N_{s}^{\mu }=0$, i.e., the torque $N_{s}$ is zero in all
relatively moving inertial frames of reference, $N_{s}=0$, when $r^{0}=0$.
In order to find the explicit expression for the torque $N_{t}$ as a CBGQ in
the $S$ frame we can simply perform the LT of the 1-vector $%
N_{t}=N_{t}^{\prime \mu }\gamma _{\mu }^{\prime }$. Then both torques $N_{s}$
and $N_{t}$ as CBGQs in the $S$ frame can be written as
\begin{eqnarray}
r^{0} &=&0,\ N_{s}=N_{s}^{\mu }\gamma _{\mu }=0,\quad N_{t}=N_{t}^{\mu
}\gamma _{\mu },  \notag \\
N_{t}^{0} &=&\gamma \beta N_{t}^{\prime 1},\ N_{t}^{1}=\gamma N_{t}^{\prime
1},\ N_{t}^{2}=N_{t}^{\prime 2},\ N_{t}^{3}=N_{t}^{\prime 3}=0,  \label{s3}
\end{eqnarray}
where the components $N_{t}^{\prime i}$ are given by (\ref{s2}). Obviously
it again holds that $N_{s}$ and $N_{t}$ from (\ref{s2}) are equal to $N_{s}$
and $N_{t}$ from (\ref{s3})
\begin{equation}
r^{0}=0,\ N_{s,t}=N_{s,t}^{\prime \mu }\gamma _{\mu }^{\prime }(\ref{s2}%
)=N_{s,t}^{\mu }\gamma _{\mu }(\ref{s3}).  \label{st1}
\end{equation}
It is important to remark that the decomposition of $N$ into $N_{s}$ and $%
N_{t}$, equations (\ref{nls}) and (\ref{cs}), will be fulfilled for both
representations (\ref{s2}) and (\ref{s3}). All this shows that the principle
of relativity is again naturally satisfied and there is no Trouton-Noble
paradox when the theory is formulated with 4D torques $N_{s}$ and $N_{t}$%
.\bigskip \medskip

\noindent \textbf{4.\ Comparison \ with the 3D torques} $\mathbf{T}$ \textbf{%
and }$\mathbf{T}_{t}$ \bigskip

\noindent The relations (\ref{l1}) indicate that in the frame of
``fiducial'' observers, here the $S^{\prime }$ frame, and in the $\{\gamma
_{\mu }^{\prime }\}$ basis one can make the identification
\begin{eqnarray}
v^{\prime \mu } &=&(c,0,0,0):T_{i}^{\prime }=N_{s}^{\prime
i}=(1/2c)\varepsilon ^{0jki}N_{jk}^{\prime }v_{0}^{\prime }=(1/2)\varepsilon
^{0jki}N_{jk}^{\prime },  \notag \\
T_{t,i}^{\prime } &=&N_{t}^{\prime i}=(1/c)N^{\prime 0i}v_{0}^{\prime
}=N^{\prime 0i}.  \label{idt}
\end{eqnarray}
Observe that such identification is fulfilled only for the components. The
3D vector $\mathbf{T}^{\prime }$ is $\mathbf{T}^{\prime }\mathbf{=}%
T_{1}^{\prime }\mathbf{i}^{\prime }+T_{2}^{\prime }\mathbf{j}^{\prime
}+T_{3}^{\prime }\mathbf{k}^{\prime }$. $\mathbf{T}^{\prime }$, as a
geometric quantity in the 3D space, is constructed multiplying the
components $T_{i}^{\prime }$ by the unit 3D vectors $\mathbf{i}^{\prime }$, $%
\mathbf{j}^{\prime }$, $\mathbf{k}^{\prime }$. In contrast to it the 4D
vector $N_{s}$ in the $\{\gamma _{\mu }^{\prime }\}$ basis is a geometric
quantity in the 4D spacetime. It consists from the components $N_{s}^{\prime
\mu }$ and the basis $\{\gamma _{\mu }^{\prime }\}$. In this case, when $%
S^{\prime }$ is the frame of ``fiducial'' observers $N_{s}$ is given as $%
N_{s}=0\gamma _{0}^{\prime }+N_{s}^{\prime i}\gamma _{i}^{\prime }$. The
same holds for $\mathbf{T}_{t}^{\prime }$ and $N_{t}$ in the $\{\gamma _{\mu
}^{\prime }\}$ basis. The bases in the 3D space and the 4D spacetime are
completely different.

In the usual approaches in which components of the ``physical'' quantities $%
\mathbf{T}$, $\mathbf{L}$, $\mathbf{B}$ and $\mathbf{E}$ are derived from
components of the corresponding covariant quantities $N^{\mu \nu }$, $M^{\mu
\nu }$ and $F^{\mu \nu }$, see [25] and for $F^{\mu \nu }$ [26] section
11.9, the components $T_{i}$ and $T_{t,i}$ are not defined by (\ref{idt})
than they are identified with the ``space-space'' and ``time-space''
components respectively of the torque four-tensor $N^{\mu \nu }$. Thus, in
the $S^{\prime }$ frame, they are determined as
\begin{equation}
T_{i}^{\prime }=(1/2)\varepsilon _{ikl}N^{\prime kl},\quad T_{t,i}^{\prime
}=N^{\prime 0i},  \label{t2}
\end{equation}
This is completely analogous to the way in which the components of 3D
vectors $\mathbf{B}^{\prime }$ and $\mathbf{E}^{\prime }$ are identified
with the ``space-space'' and the ``time-space'' components respectively of
the covariant expression for the electromagnetic field $F^{\prime \mu \nu }$%
; $B_{i}^{\prime }=(1/2c)\varepsilon _{ikl}F^{\prime lk}$, $E_{i}^{\prime
}=F^{\prime i0}$, [26] section 11.9. Obviously the components $T_{i}^{\prime
}$ correspond to $-B_{i}^{\prime }$ and $T_{t,i}^{\prime }$ to $%
-E_{i}^{\prime }$. In these equations we use the notation in which
components of $\mathbf{T}^{\prime }$, $\mathbf{B}^{\prime }$ and $\mathbf{E}%
^{\prime }$ are written with lowered (generic) subscripts, since they are
not the spatial components of 4D quantities. This refers to the third-rank
antisymmetric $\varepsilon $ tensor too. The super- and subscripts are used
only on components of 4D quantities. Geometric quantities in the 3D space, $%
\mathbf{B}^{\prime }$ and $\mathbf{E}^{\prime }$ are also formed by the
multiplication of the components $B_{i}^{\prime }$ and $E_{i}^{\prime }$
with the unit 3D vectors $\mathbf{i}^{\prime }$, $\mathbf{j}^{\prime }$, $%
\mathbf{k}^{\prime }$. However it is important to note once again that the
covariant quantities $N^{\mu \nu }$ and $F^{\mu \nu }$ (and $M^{\mu \nu }$)
are only components (numbers) that are (implicitly) determined in Einstein's
system of coordinates. Components are frame-dependent numbers
(frame-dependent because the basis refers to a specific frame). Components
tell only part of the story, while the basis contains the rest of the
information about the considered physical quantity. These facts are
completely overlooked in all usual covariant approaches and in the above
identifications (\ref{t2}) and those for $B_{i}$, $E_{i}$ $L_{i}$ and $%
L_{t,i}$.

After this digression let us go back to the identification (\ref{idt}). It
is visible from (\ref{idt}) and (\ref{s2}) that in the case when $r^{0}=0$
all components $T_{i}^{\prime }=0$ and the components $T_{t,i}^{\prime }$
are the same as $N_{t}^{\prime i}$ from (\ref{s2}). Such result for $%
T_{i}^{\prime }$ agrees with that one from [7], but in [7] there are no
components $T_{t,i}^{\prime }$, i.e., the 3D vector $\mathbf{T}_{t}^{\prime
} $.

Now comes the fundamental difference between the approaches with 3D
quantities and our approach with 4D geometric quantities. $N_{s}$ and $N_{t}$
as CBGQs in the $S^{\prime }$ frame are determined by equation (\ref{lg}).
When $S^{\prime }$ is the frame of ``fiducial'' observers then (\ref{lg})
becomes (\ref{l1}) and in that case the identification (\ref{idt}) is
possible. In order to find $N_{s}$ and $N_{t}$ as CBGQs in the $S$ frame, in
which the capacitor is moving, we have to transform by the LT\emph{\ all}
quantities, which determine $N_{s}$ and $N_{t}$ in (\ref{lg}) or (\ref{l1}),
i.e., $N^{\prime \mu \nu }$, $v_{\mu }^{\prime }$ and $\gamma _{\nu
}^{\prime }$, from $S^{\prime }$ to $S$. This procedure yields the LT (\ref
{aa}) for the components $N_{s}^{\mu }$ and similarly for $N_{t}^{\mu }$. It
is important to note that ``fiducial'' observers are moving in $S$.
Therefore the components $v^{\mu }$ of their velocity in $S$, which are
obtained by the LT from $v^{\prime \mu }=(c,0,0,0)$, are $v^{\mu }=(\gamma
c,\gamma \beta c,0,0)$. Of course, for the whole CBGQ $v$ it holds that $%
v=v^{\prime \mu }\gamma _{\mu }^{\prime }=v^{\mu }\gamma _{\mu }$. Hence, in
order to find the components $N_{s}^{\mu }$ and $N_{t}^{\mu }$ in $S$ it is
not enough to transform only $N^{\prime \alpha \beta }$ but the components $%
v^{\prime \mu }$ as well. In $S$, as seen from (\ref{aa}), the 4D torque $%
N_{s}$ contains not only the spatial components $N_{s}^{i}$ but the temporal
component $N_{s}^{0}=\gamma \beta N_{s}^{\prime 1}$ as well. Also it follows
from (\ref{aa}) that the components $N_{s}^{\prime \mu }$ transform again to
the components $N_{s}^{\mu }$ of the same 1-vector $N_{s}$. Thus, according
to (\ref{s2}) $N_{s}^{\prime \mu }=0$ in $S^{\prime }$, but according to (%
\ref{s3}) $N_{s}^{\mu }=0$ in the $S$ frame as well. Only with such
transformations, that is, with the LT of components (\ref{aa}) and the unit
1-vectors $\gamma _{\mu }^{\prime }$ it is obtained that $%
N_{s}=N_{s}^{\prime \mu }\gamma _{\mu }^{\prime }=N_{s}^{\mu }\gamma _{\mu }$%
, and the same for $N_{t}$.

On the other hand, in order to get the transformations for the components $%
T_{i}$, (\ref{s}), one has to suppose that only the components $N^{\prime
\alpha \beta }$ and not $v^{\prime \mu }$ are transformed from $S^{\prime }$
to $S$. It can be interpreted as that the components $v^{\prime \mu
}=(c,0,0,0)$ from $S^{\prime }$ are again transformed to the same $v^{\mu
}=(c,0,0,0)$ in $S$. Hence it is supposed that the same identification as (%
\ref{idt}), or (\ref{t2}), is valid not only in the $S^{\prime }$ frame, the
frame of ``fiducial'' observers, but also in the relatively moving $S$
frame. This yields the following transformations
\begin{eqnarray}
T_{1} &=&T_{1}^{\prime },\ T_{2}=\gamma (T_{2}^{\prime }-\beta
T_{t,3}^{\prime }),\ T_{3}=\gamma (T_{3}^{\prime }+\beta T_{t,2}^{\prime }),
\notag \\
T_{t,1} &=&T_{t,1}^{\prime },\ T_{t,2}=\gamma (T_{t,2}^{\prime }+\beta
T_{3}^{\prime }),\ T_{t,3}=\gamma (T_{t,3}^{\prime }-\beta T_{2}^{\prime }).
\label{tc}
\end{eqnarray}
The transformations for $T_{i}$ from (\ref{tc}) are completely equivalent to
the transformations (\ref{s}). Note that the transformations (\ref{tc}) are
the same as the transformations for the components $N^{^{\prime }\mu \nu }$ (%
\ref{nc}). This is obvious since the identifications (\ref{t2}) say that the
components of the 3D $\mathbf{T}^{\prime }$ and $\mathbf{T}_{t}^{\prime }$
are the components of the 4D torque $N$ and according to that they transform
like the components of $N^{\prime \mu \nu }$.

Furthermore it is visible from (\ref{tc}) that the components $T_{i}$ in $S$
are expressed by the mixture of components $T_{k}^{\prime }$ of the 3D
vector $\mathbf{T}^{\prime }$ and the components $T_{t,k}^{\prime }$ of
another 3D vector $\mathbf{T}_{t}^{\prime }$ from $S^{\prime }$. This is the
reason that the components of the usual 3D torque $\mathbf{T}$ will not
vanish in the $S$ frame even if they vanish in the $S^{\prime }$ frame,
i.e., that there is the Trouton-Noble paradox in the usual approaches to
special relativity.

Let us now see what is with the transformations of bases. In [7] both $%
\mathbf{T}^{\prime }$ and $\mathbf{T}$, as geometric quantities in the 3D
space, are constructed multiplying the components $T_{i}^{\prime }$ and $%
T_{i}$, given by equations (\ref{s}), by the unit 3D vectors $\mathbf{i}%
^{\prime }$, $\mathbf{j}^{\prime }$, $\mathbf{k}^{\prime }$ and $\mathbf{i}$%
, $\mathbf{j}$, $\mathbf{k}$ respectively. The components $T_{i}$ are
determined by the transformations (\ref{tc}), or (\ref{s}), but there is no
transformation which transforms the unit 3D vectors $\mathbf{i}^{\prime }$, $%
\mathbf{j}^{\prime }$, $\mathbf{k}^{\prime }$ into the unit 3D vectors $%
\mathbf{i}$, $\mathbf{j}$, $\mathbf{k}$. Hence it is not true that the 3D
vector $\mathbf{T=}T_{1}\mathbf{i}+T_{2}\mathbf{j}+T_{3}\mathbf{k}$ is
obtained by the LT from the 3D vector $\mathbf{T}^{\prime }\mathbf{=}%
T_{1}^{\prime }\mathbf{i}^{\prime }+T_{2}^{\prime }\mathbf{j}^{\prime
}+T_{3}^{\prime }\mathbf{k}^{\prime }$. Accordingly $\mathbf{T}$ and $%
\mathbf{T}^{\prime }$ determined from (\ref{tc}), or (\ref{s}), are not the
same quantity for relatively moving inertial observers, $\mathbf{T\neq T}%
^{\prime }$. Moreover, as already said, the torque $\mathbf{T}_{t}^{\prime }$
is not mentioned in [7]. Completely different situations is with 4D
geometric quantities, the torques $N$, $N_{s}$ and $N_{t}$, which are
Lorentz invariant quantities, independent of the chosen reference frame and
the system of coordinates in it. All this together shows that although the
identification (\ref{idt}) is possible in the frame of the ``fiducial''
observers, here the $S^{\prime }$ frame, it is not possible in any other
relatively moving frame, here the $S$ frame.

Comparing (\ref{tc}) with the usual transformations for $E_{k}$ and $B_{k}$,
which are

\begin{eqnarray}
B_{1} &=&B_{1}^{\prime },\ B_{2}=\gamma (B_{2}^{\prime }-\beta E_{3}^{\prime
}/c),\ B_{3}=\gamma (B_{3}^{\prime }+\beta E_{2}^{\prime }/c),  \notag \\
E_{1} &=&E_{1}^{\prime },\ E_{2}=\gamma (E_{2}^{\prime }+\beta
cB_{3}^{\prime }),\ E_{3}=\gamma (E_{3}^{\prime }-\beta cB_{2}^{\prime }),
\label{A1}
\end{eqnarray}
equation (11.148) in [26], we see that the transformations for $T_{i}$ and $%
T_{t,i}$ are the same as (\ref{A1}). In all previous treatments of special
relativity the transformations for the components $B_{k}$ and $E_{k}$, (\ref
{A1}), are considered to be the LT of the 3D electric and magnetic fields.
The same opinion exists in connection with the transformations for $T_{i}$ (%
\ref{tc}), i.e., (\ref{s}), and the transformations for $L_{i}$, equation
(11) in [25]. However in [23, 24] (Clifford algebra formalism) and [21]
(tensor formalism with tensors as 4D geometric quantities) it is \emph{proved%
} in different manners that the transformations for the 3D vectors $\mathbf{E%
}$, $\mathbf{B}$, (\ref{A1}) and equations (11.149) in [25], are not the LT
but the ``apparent'' transformations (AT) (for the name see [27]), which do
not refer to the same 4D quantity. (Previously I often called these usual
transformations for $\mathbf{E}$ and $\mathbf{B}$ as the standard
transformations.) From the analogy between the transformations for the
components $B_{k}$ and $E_{k}$, (\ref{A1}), on the one hand and (\ref{tc})
on the other hand we can conclude that the transformations (\ref{tc}) for $%
T_{i}$ and $T_{t,i}$, i.e., (\ref{s}), are also the AT. The same holds for
the transformations for $L_{i}$ equation (11) in [25], as explained in
detail in [9].

As seen from sections 3.1 - 3.3, the 4D torques $N$, $N_{s}$ and $N_{t}$ are
determined taking that $S^{\prime }$, the rest frame of the capacitor, is
the frame of ``fiducial'' observers. In section 3 in [7], under the title
``The Trouton-Noble paradox as an electrodynamic paradox,'' Jefimenko
calculated the 3D torque $\mathbf{T}$ in the $S$ frame using direct
nonrelativistic electromagnetic calculations and not the transformations for
the components $T_{i}$. The whole calculation is exclusively made with the
3D quantities $\mathbf{E}$, $\mathbf{B}$, $\mathbf{F}$ and $\mathbf{T}$. In
our approach the case corresponding to section 3 from [7], but with the 4D
geometric quantities, would be that the $S$ frame is the frame of
``fiducial'' observers and that $r^{0}=0$. This will not be investigated
here since the analogous case with Jackson's paradox is examined in detail
in [9], where it is shown that there is no paradox in that case as well and
that the principle of relativity is naturally satisfied when the 4D
geometric quantities are used.

In section 4 in [7] an additional 3D torque is introduced, which comes from
the rate of change of the 3D angular electromagnetic field momentum, and
which balances the rate of change of the 3D angular mechanical momentum of
the moving capacitor. It is argued in [7] that such procedure resolves the
paradox with the 3D torque. In our approach with 4D geometric quantities
there is no paradox and consequently there is no need for any additional
torque.

However it is worth noting that even in the case when the electromagnetic
field is concerned one has to deal with 4D AQs and not with the usual 3D
quantities. Recently, [8], I have developed an axiomatic geometric
formulation of electromagnetism with the bivector $F$ as the primary
quantity for the whole electromagnetism. There, [8], it is shown that the
most important quantity for the momentum and energy of the electromagnetic
field is the observer independent stress-energy vector $T(n)$ (1-vector),
which is a vector-valued linear function on the tangent space at each
spacetime point $x$ describing the flow of energy-momentum through a
hypersurface with normal $n=n(x)$. From that $T(n)$ some new observer
independent expressions (AQs) are obtained for the energy density $U$
contained in an electromagnetic field, for the Poynting vector $S$, for the
momentum density $g$ and for the angular-momentum density $M$. They are all
written in terms of $F$. The most general expressions for $T(n)$, $U$, $S$, $%
g$ and $M$, but with 1-vectors $E$ and $B$ and not with $F$, can be simply
obtained inserting the decomposition of $F$, (\ref{itf}), into the relations
with $F$ from [8]. We consider that the 4D geometric quantities $T(n)$, $U$,
$S$, $g$ and $M$ have to be used in all investigations of the
electromagnetic field instead of the usual expressions with the 3D $\mathbf{E%
}$ and $\mathbf{B}$. \bigskip \bigskip

\noindent \textbf{5.}\textit{\ }\textbf{Conclusions}\bigskip

\noindent The main conclusion that can be drawn from the whole consideration
is that the relativistically correct description of physical phenomena
without any paradoxes can be achieved in the consistent way with 4D
geometric quantities as physical quantities in the 4D spacetime. On the
other hand, as seen from all previous treatments of the Trouton-Noble
paradox and from our extensive discussion, the use of 3D quantities and
their AT necessarily leads to different ambiguities and inconsistencies.

Regarding the measurements of the 4D geometric quantities we give the
following remark. In order to check the validity of the relation (\ref{enc})
(and the similar relations for $N_{s}$ and $N_{t}$), and of the physical law
$N=dM/d\tau $, the experimentalists have to measure all six independent
components of $N^{\mu \nu }$ (or equivalently of $N_{s}^{\mu }$ and $%
N_{t}^{\mu }$), and also of $M^{\mu \nu }$ ($M_{s}^{\mu }$ and $M_{t}^{\mu }$%
), in both frames $S^{\prime }$ and $S$. Obviously it is more complicated
than in the usual approaches with the 3D quantities, in which the
experimentalists measure only three components of the 3D torque $\mathbf{T}$
and three components of the 3D angular momentum $\mathbf{L}$ in both frames $%
S^{\prime }$ and $S$. However, the comparisons with different experiments
that are presented in [15] and [23, 24, 8] and in this paper explicitly show
that in the 4D spacetime the relativistically correct results refer to the
4D geometric quantities and not, as generally accepted, to the 3D
quantities. \bigskip \medskip

\noindent \textbf{References\bigskip }

\noindent \lbrack 1] Trouton F T and Noble H R 1903 \textit{Philos. Trans.
R. Soc. London }

\textit{Ser. A }\textbf{202} 165

\noindent \lbrack 2] Hayden H C 1994 \textit{Rev. Sci. Instrum.} \textbf{65}
788

\noindent \lbrack 3] von Laue M 1911 \textit{Phys. Zeits.} \textbf{12} 1008

\noindent \lbrack 4] Pauli W 1958 \textit{Theory of Relativity} (New York:
Pergamon)

\noindent \lbrack 5] Singal A K 1993 \textit{Am. J. Phys.} \textbf{61} 428

\noindent \lbrack 6] Teukolsky S A 1996 \textit{Am. J. Phys.} \textbf{64}
1104

\noindent \lbrack 7]Jefimenko O D 1999 \textit{J. Phys. A: Math. Gen.}
\textbf{32} 3755

\noindent \lbrack 8] Ivezi\'{c} T 2005 \textit{Found. Phys. Lett.} \textbf{18%
} 401

\noindent \lbrack 9] Ivezi\'{c} T 2006 \textit{physics}/0602105, to be
published in 2006 \textit{Found. }

\textit{Phys. }\textbf{36 }(10)

\noindent \lbrack 10] Hestenes D 1966 \textit{Space-Time Algebra }(New York:
Gordon and Breach);

1999 \textit{New Foundations for Classical Mechanics }2nd edn (Dordrecht:

Kluwer); 2003 \textit{Am. J Phys.} \textbf{71 }691

\noindent \lbrack 11] Doran C and Lasenby A 2003 \textit{Geometric algebra
for physicists }(Cambridge:

Cambridge University Press)

\noindent \lbrack 12] Einstein A 1905 \textit{Ann. Physik.} \textbf{17} 891
tr. by Perrett W and Jeffery

G B 1952 in \textit{The Principle of Relativity} (New York: Dover)

\noindent \lbrack 13] Ivezi\'{c} T 1999 \textit{Found. Phys. Lett.} \textbf{%
12 }507

\noindent \lbrack 14] Ivezi\'{c} T 2001 \textit{Found. Phys.} \textbf{31}
1139

\noindent \lbrack 15] Ivezi\'{c} T 2002 \textit{Found. Phys. Lett.} \textbf{%
15} 27; 2001 \textit{physics}/0103026;

2001 \textit{physics}/0101091

\noindent \lbrack 16] Jefimenko O D 1997 \textit{Retardation and Relativity}
(Star City, WV:

Electret Scientific)

\noindent \lbrack 17] Ludvigsen M 1999 \textit{General Relativity,} \textit{%
A Geometric Approach }

(Cambridge: Cambridge University Press)

\noindent \lbrack 18] Sonego S and Abramowicz M A 1998 \textit{J. Math. Phys.%
} \textbf{39} 3158

\noindent \lbrack 19] M\o ller C 1972 \textit{The Theory of Relativity} 2nd
edn (Oxford: Clarendon Press)

\noindent \lbrack 20] Ivezi\'{c} T 1999 \textit{Found. Phys. Lett.} \textbf{%
12} 105

\noindent \lbrack 21] Ivezi\'{c} T 2003 \textit{Found. Phys.} \textbf{33}
1339

\noindent \lbrack 22] Ivezi\'{c} T 2002 \textit{hep-th}/0207250v2

\noindent \lbrack 23] Ivezi\'{c} T 2005 \textit{Found. Phys. Lett.} \textbf{%
18} 301

\noindent \lbrack 24] Ivezi\'{c} T 2005 \textit{Found. Phys.} \textbf{35}
1585

\noindent \lbrack 25] Jackson J D 2004 \textit{Am. J. Phys.} \textbf{72} 1484

\noindent \lbrack 26] Jackson J D 1977 \textit{Classical Electrodynamics}
2nd edn

(New York: Wiley)

\noindent \lbrack 27] Rohrlich F 1966 \textit{Nuovo Cimento B} \textbf{45} 76

\end{document}